\documentclass[aps,twocolumn,floats,prd,nofootinbib,showpacs,showkeys]{revtex4} 
\usepackage[dvips]{graphicx} %
\usepackage{epsfig,amsmath}
\usepackage{amssymb}
\usepackage{rotate}
\usepackage{color}
\usepackage[ps2pdf=true]{hyperref}
\usepackage{bm}

\DeclareFontFamily{OT1}{pzc}{}
\DeclareFontShape{OT1}{pzc}{m}{it}%
             {<-> s * [1.10] pzcmi7t}{}
\DeclareMathAlphabet{\mathscr}{OT1}{pzc}%
                                 {m}{it}

\newcommand{\be}{\begin{equation}}
\newcommand{\ee}{\end{equation}}
\newcommand{\bea}{\begin{eqnarray}}
\newcommand{\eea}{\end{eqnarray}}
       
\newcommand{\refeq}[1]{Eq.~(\ref{eq:#1})}          
\newcommand{\refeqs}[2]{Eqs.~(\ref{eq:#1})--(\ref{eq:#2})}          
          
\newcommand{\reffig}[1]{Fig.~\ref{fig:#1}}

\newcommand{\vs}{\nonumber\\}       
\newcommand{\refsec}[1]{\S~\ref{sec:#1}}          
\newcommand{\refapp}[1]{App.~\ref{app:#1}}          
 
\renewcommand{\v}[1]{\mathbf{#1}}

\newcommand{\vn}{\bm{\nabla}}
\newcommand{\vx}{\v{x}}
\newcommand{\vy}{\v{y}}
\newcommand{\vu}{\v{u}}
\renewcommand{\vr}{\v{r}}

\newcommand{\vk}{\v{k}}

\newcommand{\fNL}{f_{\rm NL}}
\newcommand{\perm}{\mbox{perm.}}

\newcommand{\<}{\langle}
\renewcommand{\>}{\rangle}

\newcommand{\vh}[1]{\bm{\hat{#1}}}

\renewcommand{\d}{\delta}
\renewcommand{\a}{\alpha}
\newcommand{\D}{\Delta}

\newcommand{\rhob}{\overline{\rho}}
\renewcommand{\H}{{\rm H}}

\newcommand{\Mpch}{\,{\rm Mpc}/h}
\newcommand{\iMpch}{\,h/{\rm Mpc}}

\newcommand{\Msunh}{\,M_{\odot}/h}

\newcommand{\Om}{\Omega_m}
\newcommand{\Ob}{\Omega_b}
\newcommand{\OL}{\Omega_\Lambda}

\renewcommand{\L}{\Lambda}
\newcommand{\p}{\mathscr{p}}
\newcommand{\M}{\mathcal{M}}

\newcommand{\s}{\sigma}

\newcommand{\intd}{\int {\rm d}}

\newcommand{\npk}{n_{\rm pk}}
\newcommand{\ubr}[1]{\bar{u}_{r #1}}
\newcommand{\rhat}{\hat{r}}
\newcommand{\vrhat}{\v{\hat{r}}}
\newcommand{\Pdu}{{\mathcal P}_{\d u}}
\renewcommand{\P}{{\mathcal P}}

\newcommand{\wbar}[2]{w^{(#1)}_{R,\,#2}}
\newcommand{\xib}{\xi_R}
\newcommand{\xir}[1]{\xi^{#1}_{R}}
\newcommand{\xirr}[2]{\xi^{#1}_{R,\,#2}}

\begin{document}

\title{Large-scale Velocities and Primordial Non-Gaussianity}

\author{Fabian Schmidt}
\affiliation{Theoretical Astrophysics, California Institute of Technology M/C 350-17,
Pasadena, CA~91125, USA}

\begin{abstract}
We study the peculiar velocities of density peaks in the presence of
primordial non-Gaussianity.  Rare, high density peaks in the initial density
field can be identified with tracers such as galaxies and clusters in the 
evolved matter distribution.  
The distribution of relative velocities of peaks is derived in the large-scale 
limit using two different 
approaches based on a local biasing scheme.  Both approaches agree, and 
show that halos still stream with the dark matter locally as well as statistically, i.e. they do
not acquire a velocity bias.  Nonetheless, even a moderate degree
of (not necessarily local) non-Gaussianity induces a significant 
skewness ($\sim 0.1-0.2$)
in the relative velocity distribution, making it a potentially interesting 
probe of non-Gaussianity
on intermediate to large scales.  We also study two-point correlations
in redshift-space.
The well-known Kaiser formula is still a good approximation on large scales,
if the Gaussian halo bias is replaced with its (scale-dependent)
non-Gaussian generalization.  However, there are additional terms not
encompassed by this simple formula which become relevant on smaller scales
($k \gtrsim 0.01\iMpch$).  Depending on the allowed level of non-Gaussianity,
these could be of relevance for future large spectroscopic surveys.  
\end{abstract}


\date{\today}

\maketitle

\section{Introduction}
\label{sec:intro}

Observations of the large scale structure in the Universe that use galaxies,
clusters or other tracers of the density field are done in redshift space:  
the distance is generally inferred using the redshift $z$, which receives
a contribution from the line-of-sight velocity of the object.  These velocities
are due to the gravitational field which is correlated with the
density field itself.  On large scales where linear perturbation theory
in the density field applies, the leading contribution
is the squashing (or stretching, in case of an underdensity) of a 
volume element in redshift space relative to real space.  In this limit,
there is a simple relation between the real- and redshift-space power
spectra, $P_g$ and $P_{g,s}$, respectively, of a tracer `$g$' \cite{Kaiser,Fisher95}:
\be
P_{g,s}(k, \mu) = \left ( 1 + \frac{f}{b_1} \mu^2\right)^2 P_g(k),
\label{eq:Kaiser}
\ee
where $\mu$ is the cosine of the $\vk$ vector with the line of sight, 
$f=d\ln D/d\ln a$ is the logarithmic derivative of the linear growth 
factor, and $b_1$ is the linear bias of the tracer population.  Apart
from the large-scale, small-correlation limit, the
relation \refeq{Kaiser} makes two assumptions: first, that the tracer 
population is characterized by a deterministic local bias.  In particular, 
if the tracer density $\d_g(\vx)$ is a local function $F(\d(\vx))$ of the 
matter density perturbation $\d$, we can expand in $\d(\vx)$ to obtain
\cite{FryGaztanaga}
\be
\d_g(\vx) = b_1\:\d(\vx) + \frac{b_2}{2} \d^2(\vx) +\dots,
\label{eq:local}
\ee  
where the bias parameters are either to be seen as free empirical parameters,
or can be determined using various theoretical approaches.  Local biasing
amounts to the assumption that dark matter halos form in high-density regions
(peaks) in the initial density field.  This assumption holds well in the
high-peak / massive halo regime, which we assume throughout.  Hence, in the
following we will somewhat loosely use ``peaks'' and ``halos'' interchangeably.

The second 
assumption used for \refeq{Kaiser} is that the cosmological density field is Gaussian on large 
scales.  While we will retain the local biasing scheme, we are interested
in relaxing the second assumption of Gaussianity.  
Recently, there has been renewed interest in probing the Gaussianity
of the primordial seed perturbations via large scale structure 
(see \cite{DesjacquesSeljak10} for a review).  
The simplest way of obtaining a non-Gaussian field is by
adding a \emph{local} non-linearity:
\be
\Phi(\vx) = \Phi_{\rm G}(\vx) + \fNL (\Phi_{\rm G}^2(\vx) - \<\Phi_{\rm G}^2\>),
\label{eq:philocal}
\ee
where $\Phi_{\rm G}$ is a Gaussian random field, and $\Phi$ is
the resulting non-Gaussian field.  
Of course, one can add higher powers
to the series \refeq{philocal}, though the quadratic term usually has the largest impact.  
Following standard convention, we
let $\Phi$ stand for the primordial potential, related to the density
field through the transfer function and Poisson equation (see \refapp{bispectra}).  

As shown by \cite{DalalEtal08} and confirmed by \cite{MV08,SlosarEtal},  in
the presence of non-Gaussian initial conditions of the local type,
halos acquire a scale-dependent correction to their bias which becomes
important on large scales:
\be
b_1 \rightarrow b_1 + 2\fNL (b_1-1)\d_c \M(k),
\ee
where $\M(k) \propto k^{-2}$ is the relation between density and potential
in Fourier space (see \refapp{bispectra}).  

Besides the local model, several other possible bispectrum configurations 
have been proposed in the literature, such as the 
\emph{equilateral} (e.g., \cite{CreminelliEtal06}) and 
\emph{folded} types (e.g., \cite{MeerburgEtal09}).   We summarize these types 
of non-Gaussianity and the relation between the potential and matter 
perturbations in \refapp{bispectra}.

Given the significant impact of (local) non-Gaussianity on the power spectrum
of biased tracers, it is then natural to ask what happens to the power
spectrum in redshift space, and whether \refeq{Kaiser} still holds.  
Furthermore, the distribution of relative velocities between tracers at
$\vx_1$ and $\vx_2$,
\be
\d \vu = \vu(\vx_2)-\vu(\vx_1)
\ee
is itself an interesting probe of non-Gaussianity \cite{CatelanScherrer}.

In this paper, we assume sub-horizon scales throughout, and adopt the 
Newtonian gauge;  further, we work in Lagrangian coordinates.  At first 
order, the
transition to Eulerian coordinates (which all observations as well as
simulation measurements are made in) simply amounts to replacing the
Lagrangian bias parameters with their Eulerian counterparts, in case
of the linear bias simply $b_{E,1} = b_{L,1}+1$.  

While we will focus on the local type of primordial non-Gaussianity for the
most part, the expressions obtained can easily be evaluated for any given 
primordial bispectrum, and we will show selected results for other bispectrum
shapes.

The paper is structured as follows: \refsec{not} introduces density
and velocity correlations, and some notation.  \refsec{deriv} contains
two different derivations of the moments of the relative peak velocity 
distribution.  
We discuss the distribution of peak velocities in \refsec{Pdu}.  
Finally, \refsec{Pks} presents the power spectrum of peaks (halos) in 
redshift space.  We conclude in \refsec{concl}.

\section{Preliminaries: Density and velocity fields}
\label{sec:not}

We consider the \emph{linear} overdensity and velocity fields of matter
$\d(\vx), \v{v}(\vx)$  as random fields, related by the (linear) continuity 
equation,
\be
\dot{\d} + \vn\v{v} = 0.
\ee
Since for the linear density field $\dot{\d} = aH f\,\d$,
we can define a scaled velocity $\vu$ which satisfies
\be
\vu\equiv \frac{\v{v}}{a H f} \Rightarrow \vn \vu = -\d.
\ee
In the following, we will always deal with smoothed density and velocity
fields which are indicated by a subscript $R$:
\bea
\d_R(\vx) &=& \intd^3\vy\: W_R(|\vx-\vy|)\:\d(\vx) \\
\vu_R(\vx) &=& \intd^3\vy\: W_R(|\vx-\vy|)\:\vu(\vx),
\eea
where $W_R$ is a normalized window function.  Specifically, we use a
real-space tophat filter for $W_R$, though the shape of the window 
function has negligible impact on our results.  Following the standard
convention, we choose the smoothing scale $R$ corresponding to a halo of 
mass $M$ to be determined by $R = (3 M /4\pi \rhob)^{1/3}$, where $\rhob$ is the 
background matter density.  

Below, we will need two-point
correlations of the density, velocity, and the cross-correlation between
the two.  Using the smoothed matter power spectrum,
\be
P_R(k) = \tilde{W}^2_R(k)\:P(k)
\ee
where $P(k)$ is the unsmoothed power spectrum and $\tilde{W}$ is the 
Fourier transform of the window function, the two-point correlations
are given by:
\bea
\xib = \xir{\d\d}(r) &\equiv& \<\d_R(\vx_1) \d_R(\vx_2)\>\\
&=& \frac{1}{2\pi^2}\intd k\:k^2 P_R(k) j_0(kr)\vs
\xir{\d u}(r) &\equiv& \< \d_R(\vx_1)\: \vrhat\cdot\vu_R(\vx_2)\>\\
&=& -\frac{1}{2\pi^2}\intd k\:k P_R(k) j_1(kr)\vs
\xir{u u}(r) &\equiv& \frac{1}{3} \< \vu_R(\vx_1) \cdot\vu_R(\vx_2)\>\\
&=& \frac{1}{6\pi^2}\intd k\: P_R(k) j_0(kr).\nonumber
\eea
Here, $r=|\vx_2-\vx_1|$ and $\vrhat = (\vx_2-\vx_1)/r$.  
$\xir{\d u}(r)$ denotes the cross-correlation between the density
at point $\vx_1$ and the matter velocity at point $\vr_{2}$ projected along
the separation vector.  It is negative, since in the presence of an overdensity 
$\d_R(\vx_1) > 0$ the streaming motion will
be directed towards point $\vx_1$.  Finally, we define the variance of the smoothed density field $\s_R$
and the one-dimensional smoothed velocity dispersion $\s_u$:
\be
\s^2_R \equiv \xir{\d\d}(0);\quad \s_u^2 \equiv \xir{u u}(0).
\ee
We will also need expressions for various three-point correlators
of the line-of-sight velocity and density.  The density three-point 
function is given in terms of the smoothed matter bispectrum $B_R$,
which is defined in \refapp{bispectra}, by
\begin{align}
& \xir{\d\d\d}(\vx_1,\vx_2,\vx_3) = \int \frac{{\rm d}^3\vk_1}{(2\pi)^3}
\int \frac{{\rm d}^3\vk_2}{(2\pi)^3} \\
& \  \  \times\;e^{i[\vk_1\cdot(\vx_1-\vx_3) + \vk_2\cdot(\vx_2-\vx_3)]} 
\; B_R(\vk_1,\vk_2,-\vk_1-\vk_2).\nonumber
\end{align}
Since we are interested in two-point peak correlations, we will always
encounter degenerate triangles with $\vx_3=\vx_2$.  If instead two other
vertices coincide, we can always relabel the indices to bring the
correlation into this form.  
We define the general, mixed density/velocity degenerate three-point 
correlation as
\bea
\xir{l m n}(r) &=& \< X^{(l)}(\vx_1) X^{(m)}(\vx_1) X^{(n)}(\vx_2) \>,\label{eq:xilmn}\\
X^{(l)}(\vx) &=& \left \{ 
\begin{array}{rl}
\d_R(\vx), & l=0 \vspace*{0.1cm}\\
\vh{r}\cdot\vu_R(\vx), & l=1,
\end{array} \right .
\eea
where $\vr = \vx_2-\vx_1$.  
Using that in Fourier space, $\tilde\vu_R\cdot\vh{r} = \vh{k}\cdot\vh{r}/(-i k) 
\tilde\d_R$, and
noting our definition of $\vr$, we can write these correlations as
\bea
\xir{l m n}(r) &=& \int \frac{{\rm d}^3\vk}{(2\pi)^3} e^{i\vk\cdot\vr}
\left(\frac{\vh{k}\cdot\vh{r}}{-i k}\right)^n
\int \frac{{\rm d}^3\vk_2}{(2\pi)^3} 
\left(\frac{\vh{k}_2\cdot\vh{r}}{-i k_2}\right)^m\vs
& &\times\left(\frac{-(\vk+\vk_2)\cdot\vh{r}}{-i |\vk+\vk_2|^2}\right)^l
 B_R(\vk+\vk_2,\vk_2,\vk).
\eea
The corresponding Fourier-space quantity can be defined by
\bea
\xir{l m n}(r) &=& \int \frac{{\rm d}^3\vk}{(2\pi)^3} e^{i\vk\cdot\vr}\:
i^{l+m+n} \p^{l m n}(\vk) \vs
\p^{l m n}(k) &=& \left(\frac{1}{k}\right)^n
\int \frac{{\rm d}^3\vk_2}{(2\pi)^3} 
\left(\frac{-k_z-k_{2z}}{|\vk+\vk_2|^2}\right)^l\label{eq:plmn}\\
& & \times\;\left(\frac{k_{2z}}{k_2^2}\right)^m B_R(\vk+\vk_2,\vk_2,\vk),\nonumber
\eea
where here and throughout we use the small-angle (flat sky) approximation,
and let the line-of-sight direction be along the $z$-axis.  Further, we
have taken out powers of $i$ in order to make $\p^{lmn}$ real.  

For the numerical results, we adopt a flat $\L$CDM cosmology, 
with $\Om=1-\OL=0.28$, $\Ob=0.046$, $\s_8=0.8$, and $n_s=0.958$.  The 
Gaussian Lagrangian bias of halos is calculated as $b_L = \d_c/\s_R^2$, 
where $\d_c=1.686$ is the linear collapse threshold.  Note that
our expressions are more general in the sense that they can be used with any 
local bias parameters (see \refsec{ansatz}).  

\section{Velocities of density peaks}
\label{sec:deriv}

In this section we derive the moments of the distribution of velocities
of density peaks in two different ways, both based on the local bias model
of \refeq{local} (i.e., local in terms of the \emph{non-Gaussian} density field $\d$).  
We are interested in the
relative velocity of two peaks projected on the separation vector,
\be
\d u \equiv [\vu(\vx_2) - \vu(\vx_1)]\cdot \vrhat.
\ee
This is the relevant quantity to compare with observations of large
scale velocities, in particular redshift-space distortions (see \refsec{Pks}).  
Homogeneity and isotropy dictate that the statistical properties of $\d u$ 
such as its moments only depend on $r=|\vx_2-\vx_1|$.  

\subsection{Unbiased velocity ansatz}
\label{sec:ansatz}

The first, much simpler approach is to \emph{assume} that the velocities
of peaks are \emph{locally} and \emph{statistically} unbiased.  The absence
of a local velocity bias is motivated by the fact that local processes
such as gravitational collapse of small-scale overdensities should not
induce any relative velocity with respect to the large-scale flow of matter.  In
Gaussian N-body simulations, this ansatz has been shown to hold well on
large scales, and one would not expect that mildly non-Gaussian initial 
conditions will change this.  

On the other hand, as shown in \cite{Desjacques08,DesjacquesSheth10}, 
density peaks have
a \emph{statistical}, scale-dependent velocity bias when the 
peak constraint is employed.  Here, statistical means that 
each halo still moves with the dark matter locally (no local velocity
bias), but a velocity bias still comes about from the fact that peaks in the
density field are in special locations.  This velocity bias from
the peak constraint scales as $k^2$, and we will neglect it here for simplicity
since we are interested in large scales.  However, it is not apparent whether
a non-Gaussianity of the density field on large scales can also source such
a statistical velocity bias.  We defer this question to \refsec{MLB}, and
assume in this section that there is no such bias.

Using the ansatz of unbiased velocities, it is easily possible to derive the 
first few moments of the velocity difference distribution.  For this, we expand
the peak density to second order (in Lagrangian space):
\be
\d_{\rm pk}(\vx) = b_{L} \d_R(\vx) + \frac{1}{2} b_{L,2} \d_R^2(\vx),
\ee
where the second order bias parameter is $b_{L,2} = b_L^2$
in the high-peak  (high-significance) limit we work in (see below for the 
general result).  For the sake
of clarity, we will omit the subscript $R$ for the smoothed fields, and
let $\d_i = \d_R(\vx_i)$ and similarly for $u$ for the remainder of this
section.  

By assumption, 
$\vu_{\rm pk} = \vu$, and 
we have for the relative velocity (to first order in the correlations):
\bea
\<\d u\>(r) &=& \Bigg\< [1 +  b_L \d_1 + \frac{1}{2} b_{L,2} \d_1^2]
[1 +  b_L \d_2 + \frac{1}{2} b_{L,2} \d_2^2] \vs
& & \times\;\vrhat\cdot[\vu_2-\vu_1] \Bigg\>\vs
&=& 2 b_L \xir{\d u}(r) + 2 b_L^2 \xir{\d\d u}(\vx_1,\vx_2,\vx_2)\vs
& & + b_{L,2} \xir{\d\d u}(\vx_1,\vx_1,\vx_2)\vs
&=& 2 b_L \xir{\d u}(r) - 2 b_L^2 \xir{\d u\d}(r)
+ b_L^2 \xir{\d\d u}(r).
\label{eq:dubar}
\eea
Here we have used \refeq{xilmn} to express the various three-point
correlations in a uniform way.  Note that when reordering the arguments of 
the three-point function in order 
to bring them into the form of \refeq{xilmn}, this can entail a change of sign 
when swapping $\vx_1\leftrightarrow\vx_2$ for terms odd in $u$ (due to
the projection onto $\vrhat$).  

We see that while we recover the standard expression in the Gaussian case
(obtained by setting all three-point functions to zero; see e.g. \cite{Desjacques08}),
there are several three-point terms contributing to the mean
relative velocity.  Note however that $\<\d u\>$ is non-zero only
for biased tracers $b_L\neq 0$ even in the presence of non-Gaussianity.  

Next, the dispersion of velocity differences is given by
\bea
\<\d u^2\>(r) &=& \left\< [1 +  b_L \d_1]
[1 +  b_L \d_2] (\vrhat\cdot[\vu_2-\vu_1])^2 \right\>\vs
&=& 2 \s_u^2 + 2\xir{uu}(r) + 2 b_L \<\d_1 (\vrhat\cdot[\vu_2-\vu_1])^2\>\vs
&=& 2 \s_u^2 + 2\xir{uu}(r)\vs
& & + 2 b_L [\xir{uu\d}(0) + \xir{uu\d}(r) - 2\xir{\d uu}(r) ].\label{eq:sigmadu}
\eea
Again, we recover the standard expression in the Gaussian case, with several 
additional three-point contributions.  These contributions vanish however
for an unbiased tracer ($b_L=0$).  

It is straightforward to generalize $\<\d u\>$ and $\<\d u^2\>$ to the case
when two different tracers with arbitrary linear and quadratic bias parameters
are considered.  
The first two moments of the relative velocity distribution between different
tracers are then given by:
\begin{align}
\<\d u\>(r) =\;& (b_{1a}+b_{1b})\, \xir{\d u}(r) - 2 b_{1a} b_{1b}\, \xir{\d u\d}(r)\\
 & + \frac{1}{2} (b_{2a} + b_{2b})\, \xir{\d\d u}(r).\vs
\<\d u^2\>(r) =\;& 2 \s_u^2 + 2\xir{uu}(r)\\
 & + (b_{1a}+b_{1b})\, [\xir{uu\d}(0) + \xir{uu\d}(r) - 2\xir{\d uu}(r) ].\nonumber
\end{align}
Here, $b_{1a},b_{1b}$ denote the linear (Lagrangian or Eulerian) bias 
parameters for the two tracers, while $b_{2a},b_{2b}$ denote the corresponding 
quadratic biases [\refeq{local}].  

Finally, there is also a third moment of $\d u$, resulting in a skewed 
velocity difference distribution as already noticed by \cite{CatelanScherrer}.
 It is given by
\bea
\< \d u^3\>(r) &=& \left\< [1 +  b_L \d_1]
[1 +  b_L \d_2] (\vrhat\cdot[\vu_2-\vu_1])^3 \right\>\vs
&=& \< [\vrhat\cdot(\vu_2-\vu_1)]^3\> = 6\: \xir{uuu}(r),
\label{eq:skewdu}
\eea
where for the second equality we have again assumed the absence of any
four- and higher-point terms.  
The third moment of velocities is obviously absent in the Gaussian case, 
but for non-Gaussian
initial conditions it is non-zero even for unbiased tracers (e.g., matter itself).  
In the next section we outline the second derivation of these results
(more details can be found in \refapp{deriv}).

\subsection{Derivation in statistical field theory}
\label{sec:MLB}

The assumption that peak velocities are still statistically unbiased in the
presence of large-scale non-Gaussianity seems natural.  On the other hand,
identifying peaks in the density field is a non-linear process, and thus it is 
desirable to have a proof for this assumption.  
Our second derivation of the distribution of relative velocities provides
such a proof without making any assumptions apart from the large-scale limit 
(i.e., small correlations), and local biasing in the physical, non-Gaussian 
density field.  

In accordance with local biasing (see e.g. \cite{MLB,FryGaztanaga}), 
we associate peaks with positions where the smoothed density field $\d_R$ is
larger than a threshold $\nu \s_R$ (hence, for $\nu=3$ we are selecting $3\s$ peaks
of the smoothed density field).  
The average number density of such peaks is then given by
\be
\npk = \< \Theta[\d_R(\vx) - \nu \s_R] \>,
\label{eq:npk}
\ee
where $\Theta$ is the Heavyside function, and $\<\cdot\>$ here denotes the 
\emph{ensemble} average. 
Similarly, the correlation function of peaks is given by (using the
same shorthand notation as above)
\be
\xi_{\rm pk}(r) = \frac{\< \Theta[\d_1 - \nu \s_R] \Theta[\d_2 - \nu \s_R] \>}{\npk^2} - 1.
\label{eq:xipk}
\ee
We are interested in the pairwise velocity of peaks along the separation
$r$.  
We can write the pair-weighted probability distribution for the two
velocities $\ubr{1}$, $\ubr{2}$ as:
\begin{align}
\P(\ubr{1},\ubr{2};\vx_1,\vx_2) = \npk^{-2} \;\times & \label{eq:Pv}\\
\big\< \d_D(\vrhat\cdot\vu_1 - \ubr{1})
\d_D(\vrhat\cdot\vu_2 - \ubr{2}) 
&\Theta[\d_1 - \nu \s_R] \Theta[\d_2 - \nu \s_R] \big\>.\nonumber
\end{align}
Note that by construction, the peak correlation function \refeq{xipk} is
recovered through
\be
1 + \xi_{\rm pk}(r) = \intd \ubr{1}\intd\ubr{2} \P(\ubr{2},\ubr{2}; \vx_1,\vx_2).
\label{eq:xipk2}
\ee
Given \refeq{Pv}, the volume-weighted distribution $\Pdu$ of the pairwise 
velocity $\d u \equiv \ubr{2}-\ubr{1}$
at a given separation is obtained as
\be
\Pdu(u; \vx_1,\vx_2) = \frac{1}{1+\xi_{\rm pk}(r)}\intd \ubr{1} \P(\ubr{1}, u + \ubr{1}; \vx_1,\vx_2).
\label{eq:Pu}
\ee
We now write the ensemble average of any functional ${\mathcal F}[\vu]$ of the
velocity field as a functional integral over all possible
realizations of the (linear) velocity field $\vu(\vx)$:
\be
\< {\mathcal F}[\vu] \> = \int {\mathcal D}[\vu(\vx)] {\mathcal P}[\vu] {\mathcal F}[\vu]
\ee
We make use of a result of statistical field theory (see e.g. 
\cite{GrinsteinWise,MLB,Matsubara03} for related applications), expressing the
partition function $Z[\v{J}]$ in terms of higher order (connected) correlation 
functions:
\bea
Z[\v{J}] &=& \int {\mathcal D}[\vu(\vx)] {\mathcal P}[\vu] \exp\left\{ i 
\intd^3\vx\; \v{J}(\vx)\cdot\vu(\vx) \right\}\quad\  \label{eq:Z1}\\
&=& \exp\left\{\sum_{n=2}^{\infty}\frac{i^n}{n!} \intd^3\vx_1\dots\intd^3\vx_n\right.\vs
& &\times\; \xi^{(n)}_{i_1\dots i_n}(\vx_1,...\vx_n) J^{i_1}(\vx_1)\dots J^{i_n}(\vx_n) \Bigg\}.
\label{eq:Z2}
\eea
Here,
\be
\xi^{(n)}_{i_1\dots i_n}(\vx_1,\dots\vx_n) = 
\<u_{i_1}(\vx_1)\dots u_{i_n}(\vx_n)\>_{\rm con}
\ee
is the connected $n$-point correlation function of the \emph{unsmoothed} velocity field.  
We can now use $\d_R = -\vn\vu_R$ together with \refeq{Z1} to obtain the 
velocity probability distribution of peaks defined in \refeq{Pv}, in close analogy to the derivation
for density correlations presented in \cite{MLB}.  This calculation is
detailed in \refapp{deriv}.  

We only keep terms 
linear in the correlations, an approximation valid on large scales.  
Once we have the expression for $\Pdu$ [\refeq{Pu}], it
is straightforward to obtain the moments of the relative peak velocity
$\d u$ via
\be
\<\d u^n\>(r) = \intd \d u\: \d u^n \Pdu(\d u; r).
\ee
Note that the central and non-central moments agree to first order in
the correlations (since $\<\d u\>$ is already first order).  We then
obtain (\refapp{deriv}):
\begin{align}
\<\d u\>  &= 2 b_L \xir{\d u}(r) \label{eq:ubar}\\
&\quad\   + b_L^2 [\xir{\d\d u}(r) - 2 \xir{\d u \d}(r)]\vs
\< \d u^2 \> &= 2[\s_u^2 + \xir{uu}(r) + b_L \xir{uu\d}(0)\\
&\quad\   + b_L \xir{uu\d}(r) - 2 b_L\xir{\d uu}(r)]\vs
\< \d u^3 \> &= 6\:\xir{uuu}(r)\label{eq:uskew} \\
\< \d u^4 \>  - 3 \< \d u^2 \>^2 &= 0.
\end{align}
These moments agree with those derived in \refsec{ansatz}.  
Note that there is no connected fourth moment (kurtosis) at this order, 
since we have assumed no primordial trispectrum.  The fact that the moments
of the peak velocity distribution agree with those derived in \refsec{ansatz}
shows that no \emph{large scale} statistical velocity bias of peaks is induced by
non-Gaussianity.  

\begin{figure}[t!]
\centering
\includegraphics[width=0.45\textwidth]{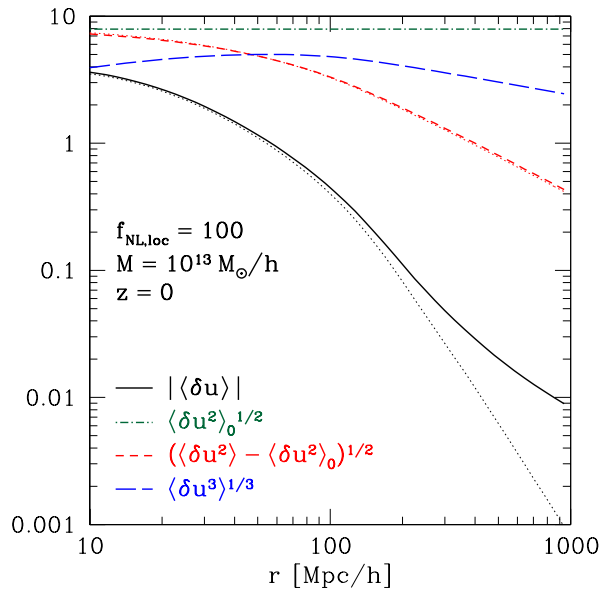}
\caption{Moments of the relative velocity distribution $\Pdu$ as function of 
the separation $r$.  The thick lines show the moments for a cosmology with
primordial non-Gaussianity of the local type ($\fNL=100$), while the thin 
dotted lines show those for vanishing non-Gaussianity.  We have separated
the variance into a scale-independent part $\<\d u^2\>_0$ and a
scale-dependent part (see text).  In all cases we assumed
halos with $M=10^{13}\Msunh$ at $z=0$ (smoothing scale $R\approx 3.1\Mpch$,
$b_L=0.88$).
\label{fig:moments}}
\end{figure}

\section{Distribution of relative velocities}
\label{sec:Pdu}

\reffig{moments} shows the moments of $\d u$ in the Gaussian 
and non-Gaussian case as function of the separation $r$.  Here, we have assumed local non-Gaussianity with
$\fNL=100$, and halos with $10^{13}\Msunh$ at $z=0$, corresponding to
a smoothing scale of $R=3.1\Mpch$ and a Lagrangian bias of $b_L=0.88$.  

The mean streaming velocity of halos $\<\d u\>$ is always negative, since 
overdense regions have a net infall velocity.  It is clearly enhanced on 
large scales by non-Gaussianity.  This is mainly due to the term 
$b_L^2 \xir{\d\d u}$, which corresponds to the density-weighted analog
of $\xir{\d u}$ and is negative for positive $\fNL$, i.e. for a positively
skewed density field.  
The effect on the variance $\<\d u^2\>$ on the other hand is very small.  
In \reffig{moments}, we have split the variance into a scale-dependent and
scale-independent part, and denoted the latter by $\<\d u^2\>_0$.  The
scale-independent standard deviation of $\d u$ is $7.9\Mpch$ for the adopted
smoothing scale and $\fNL=100$, while it is $8.1\Mpch$ in the Gaussian case, a difference
of only 3\%.  Thus, the effect of non-Gaussianity on the scale-independent
as well as scale-dependent part (\reffig{moments}) of the variance is
small.  Note that the non-Gaussian terms in 
the variance are proportional to the bias $b_L$.  

Finally, the third moment of $\d u$ is significant.  The skewness, defined
as $\<\d u^3\>/\<\d u^2\>^{3/2}$, is between 0.1 and 0.2 on a wide range
of scales.  The skewness which quantifies how strongly non-Gaussian the
velocity distribution is, depends only weakly on the halo mass, through
the smoothing scale $R$, and redshift.  
Since $\<\d u^2\>$ slowly declines with increasing smoothing scale while
$\<\d u^3\>$ is essentially independent of $R$, the 
skewness is slightly larger for higher mass halos, e.g. 10\% higher for
$10^{14}\Msunh$ halos compared to $10^{13}\Msunh$.  Note that for typical 
choices of the smoothing scale, there will be a significant non-linear 
correction to the variance $\<\d u^2\>$.  

Further, it is important to note that the skewness is non-zero for other 
bispectrum shapes as well.  For example, a bispectrum of the equilateral type
(see \refapp{bispectra}), again with $\fNL=100$, yields a skewness of
about 0.06 at $r=100\Mpch$, only a factor of three smaller than the skewness
obtained in the local model.  On the other hand, the power spectrum of halos 
only receives a scale-independent correction in the equilateral model
(\cite{VM09}, \refsec{Pks}).  Thus, the distribution of relative velocities
could be an interesting avenue to test especially those models of non-Gaussianity
which can only be weakly constrained from the clustering of halos.  

Generally, the skewness is positive for positive $\xir{uuu}(r)$,
which is usually the case when $\xir{\d\d\d}$ is positive (i.e., for
positive $\fNL$).  At first this might seem
counterintuitive, since for a positively skewed density field there
are more overdense than underdense regions.  Hence, one expects that 
velocities projected on $\rhat$ are preferentially ``inward'', i.e. negative. 
However, this is only true for density-weighted velocity correlations, such
as $\xir{\d\d u}$ which is indeed negative.  $\xir{uuu} = \<u_{1r}^2 u_{2r}\>$ 
on the other hand is a \emph{velocity-weighted} velocity correlation.  
$\xir{uuu} > 0$ thus means that high velocity regions on average move apart
(while there is no such net motion in the Gaussian case).  
High-velocity regions correspond to infall regions of overdensities, so
$\xir{uuu}$ corresponds to enhanced infall motions.  
This is not unexpected for a density field with more high-density
regions.  Of course, the converse holds for a negatively skewed density
field ($\fNL < 0$). 

\begin{figure}[t!]
\centering
\includegraphics[width=0.45\textwidth]{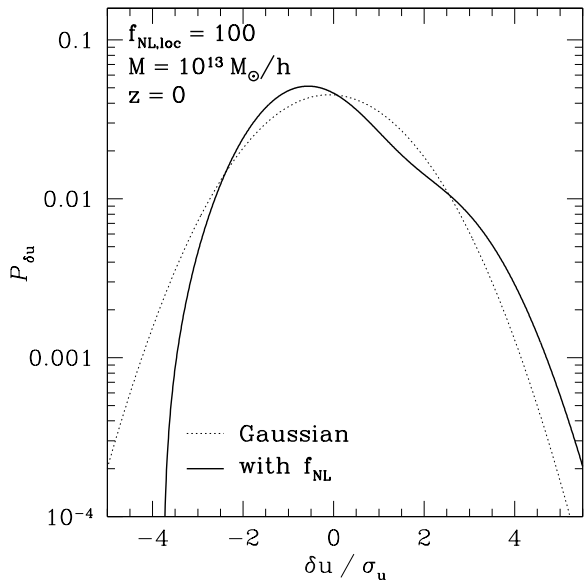}
\caption{Distribution of the relative velocity $\delta u$ for a separation
$r=100\Mpch$ in the non-Gaussian (solid) and Gaussian (dotted) case, for
the same parameters and halos as in \reffig{moments}.  Note that an equilateral
model with $\fNL\approx 300$ would lead to a result very similar to the local
model shown.
\label{fig:Pu}}
\end{figure}

It is also interesting to look at the full distribution of $\Pdu(\d u)$.  
Since by assumption the fourth and higher velocity moments are small (and in fact all
moments other than the variance are small on large scales), we can employ
the Edgeworth expansion (e.g., \cite{ScherrerBertschinger,Matsubara03}):
\bea
\Pdu(\d u;\vx_1,\vx_2) &=& \left\{ 1 + \frac{\<\d u\> \d u}{2\s_u^2} \right .
+ \frac{\<\d u^2\>}{8\s_u^2}\left(\frac{\d u^2}{\s_u^2}-2 \right) \vs
& & \  \  \left . +\frac{\<\d u^3\>}{12 \s_u^3} \frac{\d u}{4\s_u} \left(\frac{\d u^2}{\s_u^2} - 6\right)  \right\} \vs
& & \times \frac{1}{\sqrt{4\pi \:\s_u^2}} \exp\left(-\frac{\d u^2}{4\s_u^2}\right ).
\label{eq:Pdu-mom}
\eea
We also arrive at this expression directly using the statistical approach 
of \refapp{deriv}.  The distribution of the relative velocity $\d u$ 
at a separation of $100\Mpch$ is shown in \reffig{Pu} in
the Gaussian and non-Gaussian case, for the same halos and local $\fNL=100$ as in 
\reffig{moments}.  The effect of non-Gaussianity is clearly visible as
skewing the distribution towards positive velocities, as expected from
the positive third moment.  Correspondingly,
a negative $\fNL$ would lead to a velocity distribution skewed towards
negative relative velocities.  Again, these results are only weakly dependent
on halo mass and redshift.  Furthermore, the relative velocity distribution of 
the same halos in an equilateral model with $\fNL \approx 300$ looks
very similar to the local model result shown in \reffig{Pu}.  

Recently, observational studies have found evidence for larger bulk flow
motions than expected in $\L$CDM.  These studies have used galaxy peculiar
velocity surveys \cite{FeldmanEtal}, the kinetic Sunyaev-Zeldovich (SZ)
effect \cite{KashlinskyEtal}, and the estimated initial relative velocity
of the merging halos in the Bullet cluster \cite{LeeKomatsu}.  
\reffig{Pu} shows that even a moderate primordial non-Gaussianity that
is still allowed at the 2$\sigma$ level can significantly alter the tails
of the velocity distribution.  Thus an obvious question is whether
the existence of primordial non-Gaussianity can alleviate the tension
between the reported bulk flow observations and the $\L$CDM scenario.  Unfortunately,
it is not straightforward to relate our predictions of linear perturbation 
theory in the large-scale limit to these observational 
results.  Thus, we do not attempt any quantitative comparison here.  

\section{Matter and Peak Correlations in Redshift Space}
\label{sec:Pks}

One of the most important observational effects of large-scale velocities is 
their impact on correlations of tracers such as galaxies, clusters, or the
Lyman-$\alpha$ forest, through redshift distortions.  Positions are measured 
in redshift space, whose
coordinates $\v{s}$ are related to real-space coordinates $\vx$
by a shift along the line-of-sight direction:
\be
\v{s} = \vx + f u_z \v{\hat{z}}.
\label{eq:rs}
\ee
Here, we have chosen the line-of-sight to be along the $z$-axis.  Throughout,
we will work in the flat sky limit.  The statistics of the density field
(of matter or peaks) in redshift space can be determined using the real-space
statistics of the density and velocity fields together with the mapping \refeq{rs}.  

We start with the normalized distribution $\Pdu(\d u, r)$.  
The correlation function in redshift space $\xi_s$ can then be written 
as a convolution by the velocity-difference distribution $\Pdu(\d u; r)$ 
as \cite{Fisher95,Sc04}
\bea
1 + \xi_s(s_{z},\v{s}_\perp) &=& \intd \d u\; [1 + \xi(r(\d u))] \Pdu(\d u;r),\  \  \;\\
r(\d u) &=& \sqrt{\v{s}_\perp^2 + (s_z - \mu f \d u)^2},\nonumber
\eea
where the factor $\mu = r_z/r$ multiplies $\d u$, the total velocity along
$r$, to obtain the velocity along the line-of-sight.  This 
expression is exact in the flat sky limit.  
On large scales, the displacements from real to redshift space
are much smaller than the separation $r$, and we can expand $\xi$ as well
as $\Pdu$ around $r_z=s_z$.  Generalizing expressions given
in e.g. \cite{Sc04}, we obtain:
\bea
1 + \xi_s(s_{z},\v{s}_\perp) &=& \sum_{n,m=0}^\infty \frac{(-\mu\,f)^{n+m}}{n!\: m!}\\
& & \times\;\frac{d^n}{ds_z^n} (1+\xi(s)) \; \frac{d^m}{ds_z^m} \< \d u^{n+m}\>(s),\nonumber
\eea
where now $\mu = s_z/s$.  
Further, we keep only those terms which are first order in large-scale correlations.  This gives
\bea
\xi_s(s_{z},\v{s}_\perp) &=& \xi(s) - f \mu\; \<\d u\>'(s) + \frac{f^2 \mu^2}{2}
\< \d u^2\>''(s)\vs
& & - \frac{f^3\mu^3}{6} \< \d u^3 \>'''(s) + \frac{f^2\mu^4}{2} \< \d u^2 \>(s) \xi''(s)\vs
& & - \frac{f^3\mu^6}{6} \< \d u^3 \>(s) \xi'''(s),\label{eq:xis}
\eea
where primes denote derivatives with respect to $s_z$.  The last two terms
are formally second order, however the moments of
the velocity $\< \d u^m \>$ can be significant even on large
scales due to small-scale motion.  However, we verified that the last, 
non-Gaussian term is negligible compared to the other terms.  Since the 
Gaussian quadratic term is quite small on large scales as well, we will neglect
those two terms in the following.

We can now Fourier transform \refeq{xis} term by term, noting that every
derivative with respect to $s_z$ brings down a power of $i k_z = i\mu k$.  
Defining the Fourier transform of the $n$th velocity moment (in the flat
sky limit) via
\be
\<\d u^n\>(r) = \int \frac{{\rm d}^3\vk}{(2\pi)^3} e^{i\vk\cdot\vr}\:
\<\widetilde{\d u^n}\>(k),
\ee
we have:
\bea
P_s(k,\mu) &=& P(k) - f i \mu^2 k \<\widetilde{\d u}\>(k) + \frac{f^2}{2} (-\mu^4 k^2)
\<\widetilde{\d u^2}(k)\>\vs
& & - \frac{f^3}{6} \mu^6 (i k)^3 \<\widetilde{\d u^3}\>(k).
\label{eq:Ps}
\eea
This expression applies to both peaks and matter.  
With the results of \refsec{deriv}, it is then straightforward to write
down the redshift-space peak power spectrum \refeq{Ps} in the presence
of non-Gaussianities.  Again, these results apply to bispectra of
arbitrary shape- and scale-dependence.

\begin{figure}[t!]
\centering
\includegraphics[width=0.45\textwidth]{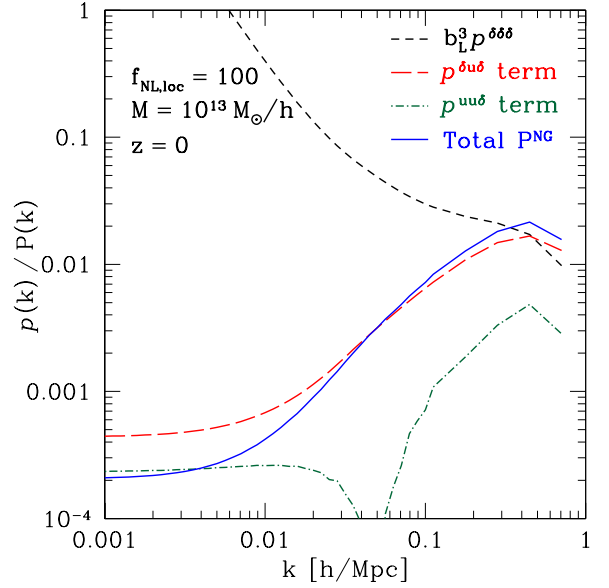}
\caption{The terms of \refeq{PNG} which contribute to the redshift-space 
power spectrum and are \emph{not} captured by the Kaiser formula, as
function of $k$.  We also show the leading real-space term
$\p^{\d\d\d}(k)$ for comparison.  All terms are for $10^{13}\Msunh$ halos at 
$z=0$ and local non-Gaussianity with $\fNL=100$.  We have set $\mu=1$ and 
divided all terms by the matter power spectrum.  Note that 
non-linear corrections to these linear predictions are expected above
$k \sim 0.05\iMpch$.  
\label{fig:terms}}
\end{figure}

Fourier transforming the $\d u$ moments given by \refeq{dubar}, \refeq{sigmadu},
and \refeq{skewdu} is easily done by replacing the correlations 
$\xi^{lmn}(r)$ with $\p^{lmn}(k)$.  
The full expression of the large-scale limit then reads
\bea
P_s(k,\mu) &=& b_L^2 P_R(k) + b_L^3 \p^{\d\d\d}\vs
& & - f \mu^2 i k [2 b_L (i k)^{-1} P_R(k) \vs
& & \qquad\qquad + b_L^2 (i \p^{\d\d u}(k) - 2i \p^{\d u\d}(k))]\vs
& & - \frac{f^2}{2}\mu^4 k^2 [ 2 (i k)^{-2} P_R(k) - 2 b_L \p^{uu\d}(k)\vs
& & \qquad\qquad\qquad + 4 b_L \p^{\d uu}(k) ]\vs
& & - \frac{f^3}{6} \mu^6 (i k)^3 6 i^3\p^{uuu}(k).
\eea
The first line contains the real-space peak power spectrum including the
well-known non-Gaussian contribution (\cite{MV08}, see \refeq{Ppk} in 
\refapp{deriv}).  Note that in Eulerian space, $b_L^3$ is to be replaced
by $b_1 b_2$.  
Using $\p^{l m u} = \p^{l m \d}/ k$, which follows from \refeq{plmn}, we get
\bea
P_s(k,\mu) &=& b_L^2 P_R(k) + 2 b_L f \mu^2 P_R(k) + f^2 \mu^4 P_R(k)\nonumber\\
& & + b_L^3 \p^{\d\d\d}(k) + f\mu^2 b_L^2 \p^{\d\d\d}(k)  \nonumber \\
& & - 2 f \mu^2 b_L^2 \: k \p^{\d u \d}(k)\vs
& & + f^2 \mu^4 b_L\: k^2  [ \p^{uu\d}(k) - 2 \p^{\d uu}(k) ]\vs
& & + f^3 \mu^6\: k^3 \p^{uuu}(k) \label{eq:Ps-f}
\eea

\begin{figure}[t!]
\centering
\includegraphics[width=0.45\textwidth]{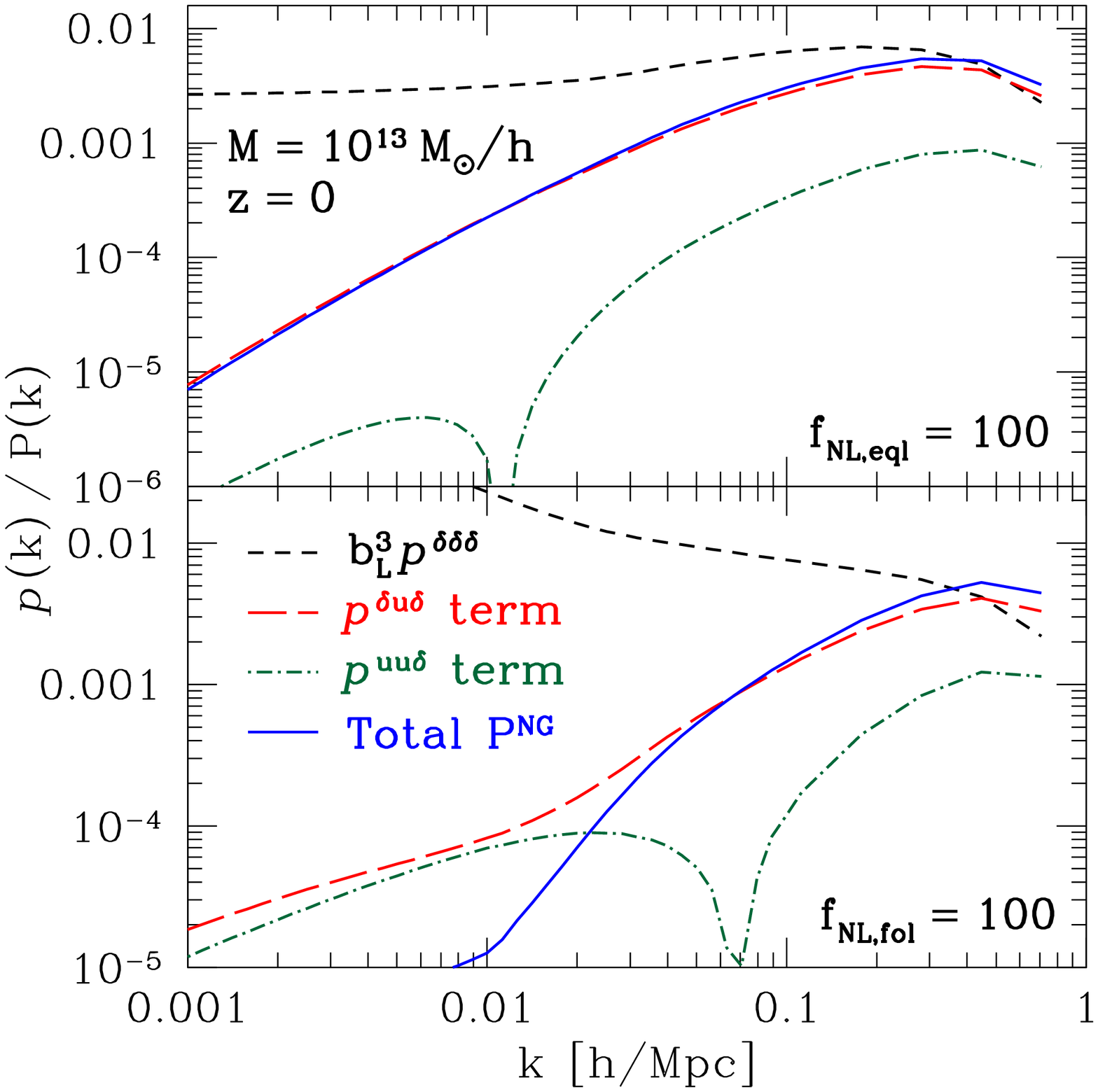}
\caption{Same as \reffig{terms}, but for equilateral (top panel) and
folded (bottom panel) primordial bispectra with $\fNL=100$ 
(see \refapp{bispectra}). 
\label{fig:terms-eql}}
\end{figure}

The terms in the first line constitute the usual Gaussian power spectrum of biased peaks
in redshift space.  Following convention \cite{DalalEtal08,MV08,SlosarEtal}, 
we can write the first 
term in the second line as a non-Gaussian correction $\D b_L$ to the
Gaussian halo bias (see e.g. \cite{MV08}):
\be
\frac{\D b_L}{b_L} = \frac{1}{2} b_L \frac{\p^{\d\d\d}(k)}{P_R(k)}.
\ee
Thus, if we include this non-Gaussian bias and neglect the terms in the last 
three lines above, the Kaiser formula \refeq{Kaiser} is still valid in the 
presence of large-scale non-Gaussianity:
\be
P^{\rm Kaiser}_s(k,\mu) = \left(1 + \frac{f \mu^2}{b_L+\D b_L}\right)^2 (b_L+\D b_L)^2 P_R(k).
\label{eq:PKaiser}
\ee
However, there are additional terms which we will denote as $P^{\rm NG}_s$:
\bea
P^{\rm NG}_s(k,\mu) &=& -2 k \p^{\d u \d}(k) \left [ f \mu^2 b_L^2 + f^2\mu^4 b_L \right ] \vs
& & + k^2 \p^{uu\d}(k) \left [ f^2\mu^4 b_L + f^3 \mu^6\right ].\;\;
\label{eq:PNG}
\eea
Corrections of order $\fNL$ to the Kaiser formula were also found by
\cite{LamDesjacquesSheth} for the variance of the density field in redshift
space.  
\reffig{terms} shows the remaining terms which violate the Kaiser formula
in the presence of non-Gaussianity on large scales.  Comparing them to
the real-space contribution $\p^{\d\d\d}$, it is clear that these terms
are far subdominant.  However, they become increasingly important towards
smaller scales.  In order to fully evaluate their importance on scales
$k \gtrsim 0.05\iMpch$, it will be necessary to extend the perturbative 
calculation to higher order or to measure the effect in N-body simulations.   
In the adopted cosmology, the non-linear scale for matter 
is $k_{\rm nl} \approx 0.27\iMpch$,
where $k_{\rm nl}$ is defined through $\D^2(k_{\rm nl}) = k_{\rm nl}^3 P(k_{\rm nl})/2\pi^2 = 1$.  Note
that $k_{\rm nl}$ will be somewhat smaller for massive halos whose power
spectrum is enhanced by $b^2$.  

\reffig{terms-eql} shows the same results for the equilateral and folded
bispectrum shapes.  The overall corrections to the Gaussian power spectrum
are much smaller, since these bispectrum shapes are suppressed in the
squeezed limit.  The relative importance of the additional terms from \refeq{PNG} is
qualitatively similar to the local case.  
Note that for all bispectra, the relative significance of these terms 
compared with the
real-space contribution $\p^{\d\d\d}$ is independent of $\fNL$ at leading 
order.

\begin{figure}[t!]
\centering
\includegraphics[width=0.45\textwidth]{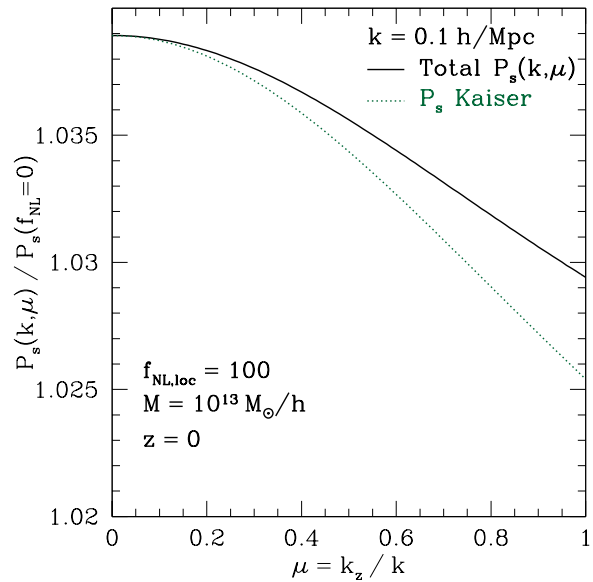}
\caption{Ratio of the redshift-space power spectrum with primordial
non-Gaussianity ($\fNL=100$) to that with Gaussian initial conditions
as function of $\mu$ at $k=0.1\iMpch$.   Again, results are shown for 
$10^{13}\Msunh$ halos at $z=0$.  Shown
is the actual ratio including all terms (\refeq{PKaiser} and \refeq{PNG}), 
and the ratio one obtains from the Kaiser formula (\refeq{PKaiser} only).  
Note that this is mainly illustrative, since we are showing linear predictions
at a scale already in the quasi-linear regime (non-linear correction from
\texttt{halofit} $\sim 5$\%).  
\label{fig:Pks}}
\end{figure}

As an example illustrating the effect on the redshift-space power spectrum, 
\reffig{Pks} shows $P_s(k,\mu)$ at $k=0.1\iMpch$ as function of $\mu$, when 
using the Kaiser formula \refeq{PKaiser} and when including all terms.  At 
these small scales, the
difference is clearly noticeable.  This is only a rough illustration however,
since in reality non-linear corrections will contribute significantly at that 
scale:  the non-linear power spectrum (from \texttt{halofit} \cite{halofit}) 
$k=0.1\iMpch$ differs by $\sim$5\% from the linear $P(k)$.  Still, we
expect that the corrections to the Kaiser formula will remain at least at the
percent-level for $k\gtrsim 0.1\iMpch$.  That would make them possibly 
relevant for upcoming spectroscopic surveys such as BOSS and HETDEX, and
certainly for future space missions such as JDEM and EUCLID, which are expected
to constrain redshift distortions at the sub-percent level \cite{McDonaldSeljak08,WhiteSongPercival}.  

Again, these results do not depend sensitively on halo mass and redshift:  
if, for example, we increase the halo mass, there is a partial cancelation 
between the larger smoothing scale and the increased halo bias, leaving the
relative size of the terms in \refeq{PNG} roughly unchanged.  For $z>0$, 
$f$ increases (asymptoting to 1 at very high redshifts), but this is
compensated by larger biases at fixed mass.

Finally, we note that the last term in \refeq{PNG}, which comes from the
skewness of the velocity distribution, also contributes to the redshift-space
power spectrum of an \emph{unbiased} tracer, for example matter itself 
(see also \cite{LamDesjacquesSheth}).  This is in 
contrast to the linear matter power spectrum in real space, which only 
receives a much smaller four-point correction of order $\fNL^2$.  

The reasons that the additional terms are so much smaller are first, that the
dominant effect on the mean streaming of halos is taken into account
by the mixed density-velocity part of the Kaiser formula (when including
the scale-dependent bias correction); and second, 
that the third moment of $\d u$, though significantly changing the distribution
of $\d u$, only affects the correlation function
via three derivatives [\refeq{xis}], or three powers of $k$ in case
of the power spectrum.  This strongly suppresses its contribution on 
large scales.  Note that this is specific to the power spectrum,
and a larger effect of $\xir{uuu}$ is expected e.g. for the bispectrum
in redshift space.

\vspace*{-0.2cm}
\section{Conclusions}
\label{sec:concl}

In this paper, we have studied the large-scale motions of peaks in the density field
in the presence of primordial non-Gaussianity.  In the high-peak regime we
are interested in, peaks can be identified with dark matter halos and visible structures
such as galaxies and clusters.  
We have derived the statistics of the relative velocity of
halo pairs in two different approaches based on a local biasing scheme.  Note
that the bias is local in the physical, \emph{non-Gaussian} density field.  
The first, simpler and more accessible approach assumes that on large scales,
halo velocities follow those of the total matter.  Given linear and
quadratic bias parameters, it is then straightforward to calculate the
moments of the relative velocity distribution.  In the second approach
(see \refapp{deriv}), no
assumptions are made apart from local biasing and the large-scale, 
small-correlation limit.  Both approaches agree, showing that the assumption
of statistically unbiased halo velocities is consistent on large scales even 
in the presence of non-Gaussianity.  

Interestingly, the presence of non-zero three-point correlations on large
scales leads to significant changes in the velocity distribution of halos,
and matter itself.  
For a positive $\fNL$, the mean streaming velocity is significantly enhanced 
on large scales, and
a non-zero third moment leads to a positive skewness of the velocity difference
distribution.  In the local model, the skewness 
$\<\d u^3\>\:/\:\<\d u^2\>^{3/2}$ is of order 
$(0.1-0.2)\times(\fNL/100)$ on a wide range of scales.  Even in the
equilateral model, which has little effect on the halo power spectrum, the 
skewness still reaches 0.06 for the same $\fNL$.  
This suggests that the velocity difference distribution can serve as an 
interesting probe of non-Gaussianity (as already pointed out by 
\cite{CatelanScherrer}).  Furthermore, these findings could be of relevance 
to recent observational reports of
significantly higher velocities than expected in the $\L$CDM (or more generally,
General Relativity + Dark Energy) scenario \cite{FeldmanEtal,KashlinskyEtal,LeeKomatsu}.  In order to evaluate this 
quantitatively however, one has to take into account non-linear corrections,
via perturbation theory and/or simulations.  

Our second result is an expression for the redshift-space power spectrum of halos (or matter)
in the presence of non-Gaussianity.  The well-known Kaiser formula relating
real- and redshift-space power spectra, extended by the scale-dependent
halo bias (see e.g. \cite{DesjacquesSeljak10}), receives non-Gaussian corrections which become relevant on small
scales.  Note that through these corrections, the redshift-space power spectrum measured at a given
scale as function of line-of-sight angle in principle allows
for a direct measurement of non-Gaussianity
from a given tracer population, without any reference to the
underlying matter power spectrum (in the Kaiser formula at a fixed scale, the
non-Gaussian effects are perfectly degenerate with the galaxy bias).  
Furthermore, the redshift-space power
spectrum of matter itself receives corrections of order $\fNL$ from 
non-Gaussianity, which are not present in real space.  

The non-Gaussian corrections to the Kaiser formula lead to the interesting
question of whether there are degeneracies of the effects of non-Gaussianity 
with other parameters measured from redshift-space 
distortions, such as dark energy parameters or consistency tests of General
Relativity \cite{ZhangEtal,SongPercival09}.  The severity of the degeneracies will depend on the level of allowed 
non-Gaussianity.  While the local model is likely to be constrained tightly
in the near future, other models such as the equilateral model are much
less constrained, but can still be lead to noticeable effects in redshift
space (\reffig{terms-eql}).  

Finally,
we expect that non-Gaussian effects on halo velocities will have an even 
larger impact on the \emph{bispectrum} of halos in redshift space.  
Again, these questions deserve more study via higher order perturbation theory 
as well as N-body simulations.

\vspace*{-0.5cm}
\begin{acknowledgments}

\vspace*{-0.2cm}
I am grateful to Vincent~Desjacques for very helpful discussions
and providing important insights in the initial stages of this project.  
Thanks also go to Uro$\check{\rm s}$~Seljak and
the Institute for Theoretical Physics at the University of Z\"urich for
their hospitality.  Further, I would like to thank Marc~Kamionkowski, 
Niayesh~Afshordi, Tobias~Baldauf, Neal~Dalal, Uro$\check{\rm s}$~Seljak, Roman~Scoccimarro, and 
especially Olivier~Dor\'e for enlightening discussions.  
This work was supported by the Gordon and Betty Moore Foundation at Caltech.  
\end{acknowledgments}

\begin{widetext}
\appendix

\section{Matter Bispectra from Primordial Non-Gaussianity}
\label{app:bispectra}

Primordial non-Gaussianity is most easily characterized by the bispectrum of 
the primordial curvature perturbations.  Different shapes, i.e. configuration 
dependences, have been proposed in the literature.  
Since we work in the sub-horizon regime at late times, 
we phrase non-Gaussianity in terms of the potential $\Phi$ at early times, 
which can be directly related to the primordial perturbations.  The 
bispectrum $B_\Phi$ at a fixed redshift is defined by
\be
\< \Phi(\vk_1)\Phi(\vk_2)\Phi(\vk_3)\> = (2\pi)^3 \d_D(\vk_1+\vk_2+\vk_3)
\: B_\Phi(k_1,k_2,k_3).
\ee
Throughout, we take $B_\Phi$ to be defined at last scattering 
(``CMB convention'', $z_*\approx 1100$).  
The bispectrum corresponding to \emph{local} non-Gaussianity of the type 
\refeq{philocal} is
\be
B^{\rm loc}_\Phi(k_1,k_2,k_3) = 2 \fNL (P_\Phi(k_1) P_\Phi(k_2) + 2\:\perm),
\ee
where $P_\Phi$ is the power spectrum of $\Phi$ at $z_*$.  The bispectra corresponding
to \emph{equilateral} and \emph{(en)folded} types are given by
\bea
B_\Phi^{\rm eql} &=& 6 \fNL \left [ -P_1 P_2 - 2\:\perm - 2 (P_1 P_2 P_3)^{2/3}
+ P_1^{1/3} P_2^{2/3} P_3 + 5\:\perm \right ]\\
B_\Phi^{\rm fol} &=& 6 \fNL \left [ P_1 P_2 + 2\:\perm + 3 (P_1 P_2 P_3)^{2/3}
- P_1^{1/3} P_2^{2/3} P_3 - 5\:\perm \right ].\\
\eea
In order to convert this to the matter bispectrum smoothed with the window 
function $W_R$, we use the Poisson equation (again, valid on sub-horizon 
scales):
\be
\d_R(\vk,z) = \frac{2}{3}\frac{k^2}{(1+z)\,H_0^2 \Om} T(k) g_*(z) \tilde W_R(k) \Phi(\vk,z_*)
\equiv \M_R(k,z) \Phi(\vk,z_*),
\ee
where $T(k)$ is the matter transfer function, and $g_*(z) \propto (1+z) D(z)$ is
the potential growth factor normalized to unity at last scattering.  
Then, power spectra and bispectra are related by
\bea
P_R(k,z) &=& \M_R^2(k,z) P_\Phi(k),\\
B_R(k_1,k_2,k_3;z) &=& \M_R(k_1,z)\M_R(k_2,z)\M_R(k_3,z) B_\Phi(k_1,k_2,k_3).
\eea

\section{Derivation of the velocity difference distribution}
\label{app:deriv}

In this appendix we present some details on the derivation of the
relative velocity distribution for peaks, starting from \refeq{Pv}.  
Expressing the Dirac and Heavyside functions in terms of their
Fourier transforms, using $\d_R = -\vn\vu_R$, and using \refeq{Z1} as well
as partial integration, we can write the velocity probability distribution
of peaks as:
\bea
\P(\ubr{1},\ubr{2};\vx_1,\vx_2) &=& \int_{\nu\s_R}d\a_1\int_{\nu\s_R} d\a_2
\left(\prod_{r=1}^4 \int \frac{{\rm d}\phi_r}{2\pi}\right) \exp(-i \a_r\phi^r)\: Z[\v{J}],\label{eq:Pu1}\\
J^i(\vx) &=& \phi_1 \partial^i W_R(|\vx-\vx_1|) + \phi_2 \partial^i W_R(|\vx-\vx_2|) \\
& & + \phi_3 W_R(|\vx-\vx_1|) \rhat^i + \phi_4 W_R(|\vx-\vx_2|) \rhat^i.
\eea
Here, we have defined $\a_3\equiv\ubr{1}$, $\a_4\equiv\ubr{2}$ in order to
allow for more compact notation, and used the Einstein summing convention.    
In the following we adopt the notation of \cite{MLB}.  Using \refeq{Z2} and extracting
the terms proportional to $\phi_r^2$, $r=1...4$,  leaves the $\phi$ integrals
as Fourier transforms of Gauss exponentials.  We obtain:
\bea
\P(\ubr{1},\ubr{2};\vx_1,\vx_2) &=& \frac{1}{(2\pi)^2 \s_u^2} \int_{\nu\s_R}d\a_1\int_{\nu\s_R} d\a_2 \label{eq:Pv2}\\
& & \exp\left\{ \frac{1}{2}\sum_{i\neq j}^{4} \wbar{2}{ij} \frac{\partial^2}{\partial\a_i\partial\a_j}
+ \sum_{n=3}^\infty \frac{(-1)^n}{n!} \sum_{[i_n]}^4\wbar{n}{i_1\dots i_n}
\frac{\partial^n}{\partial\a_{i_1}\dots\partial\a_{i_n}} \right\}
\exp\left(-\frac{1}{2} \a_r\a^r \right ).\nonumber
\eea  
Here, we have denoted ordered sequences of integers $i_1,\dots,i_n$ with $[i_n]$,
and the sum runs over all those ordered sequences with each index running
from 1 to 4.  Furthermore, we have defined
\bea
\wbar{n}{i_1\dots i_n} &\equiv& \xirr{n}{i_1\dots i_n} \s_{R,i_1}^{-1}\dots
\s_{R,i_n}^{-1}\\
\xirr{(n)}{i_1\dots i_n} &\equiv& 
\left ( \prod_{j=1}^n \intd^3\vx_j\:W_{(i_j)}^{k_j}(|\vx_j-\vr_{i_j}|) \right )
\; \xi^{(n)}_{k_1\dots k_n}(\vx_1,\dots\vx_n), \\
W_{(i)}^k(\vx) &\equiv& \left \{ 
\begin{array}{rl}
\partial^k W_R(\vx), & i=1,2\vspace*{0.1cm}\\
\rhat^k W_R(\vx), & i=3,4
\end{array} \right .\vspace*{0.2cm}\\
\s_{R,i} &\equiv& \left \{ 
\begin{array}{rl}
\s_R, & i=1,2\vspace*{0.1cm}\\
\s_u, & i=3,4
\end{array} \right .
\eea
This is a formidable array of definitions, which however allows us to
keep the treatment general.  In particular, the extension to higher-order
velocity distributions, or mixed peak density and velocity distributions
is straightforward.

In evaluating the derivatives with respect to $\a_j$, it is useful
to define the following function:
\bea
f_m(\nu) &\equiv& (-1)^m \int_{\nu\s_R}d\a \left(\frac{d}{d\a}\right)^m\exp\left(-\frac{1}{2}\a^2\right),\quad m \geq 1 \\
&=& \frac{1}{\sqrt{2^{m-1}}} \H_{m-1}(\nu/\sqrt{2}) \exp\left(-\frac{1}{2}\nu^2\right),
\eea
where $\H_m$ denotes the Hermite polynomials.  We then define the following 
coefficients:
\bea
a_m &\equiv& f_m(\nu) \approx \nu^{m-1}\exp(-\nu^2/2),\quad m \geq 1 
\label{eq:am} \\
a_0 &\equiv& \sqrt{\pi/2}\; \mbox{erfc}(\nu/\sqrt{2}) \approx \nu^{-1}\exp(-\nu^2/2)   \label{eq:a0}\\
b_m &\equiv& f_{m+1}(\ubr{1}/\s_u),\quad m \geq 0 \\
c_m &\equiv& f_{m+1}(\ubr{2}/\s_u),\quad m \geq 0.
\eea
Note that the $a_m$ differ by factors of $(2\pi)^{1/2}$ from those
defined in \cite{MLB}.  
The second approximate equality in \refeqs{am}{a0} is valid in the high-peak
limit, $\nu\gg 1$, which we will assume throughout.  Of course, we do not
make this approximation for the velocities $\ubr{1},\ubr{2}$.  The first
few velocity coefficients are
\bea
b_0 &=& \exp(\ubr{1}^2/2\s_u^2)\\
b_1 &=& \frac{\ubr{1}}{\s_u}\exp(\ubr{1}^2/2\s_u^2)\\
b_2 &=& \left(\frac{\ubr{1}^2}{\s_u^2}-1\right)\exp(\ubr{1}^2/2\s_u^2),
\eea
and correspondingly for $c_0$, $c_1$, $c_2$.  

For the peak correlation functions it is possible, by careful definition of 
different sets of integers, to
expand the exponential in \refeq{Pv2}, and reorder the terms into a sum
over powers of $a_i$ \cite{MLB}.  In our case, this expression would become
quite cumbersome however due to the different types of $\a$ parameters
for peak density and velocity.  Such an expression does not appear to be
particularly useful.  Instead, we make the assumption that
the smoothed correlation functions $\wbar{n}{}$ are much less than unity,
so that only the linear term of the expansion of the exponential needs
to be kept.  This will be appropriate in particular in our case since
we are interested in large scales where the correlations are small.  
It is then straightforward to perform the $\a$ integrals.  The end result
is that each derivative $\partial^n/\partial\a_i^n$ (including $n=0$) is 
replaced by $a_n$, if $i=1,2$, $b_n$ if $i=3$, and $c_n$ if $i=4$.  

Analogous to (but much simpler than) the above derivation, one can obtain an 
expression for the
peak density \refeq{npk} in terms of the density $n$-point functions:
\bea
\npk(\nu) &=& (2\pi)^{-1/2} \int_\nu d\a \exp\left\{\sum_{n=3}^\infty \frac{(-1)^n}{n!}
w_R^{(n)}(0,\dots,0) \frac{\partial^n}{\partial\a^n}\right\} e^{-\a^2/2}\\
&\approx& (2\pi)^{-1/2} \left(a_0 + \frac{1}{6}w_R^{(3)}(0,0,0) a_3\right)\\
&\approx& \frac{e^{-\nu^2/2}}{\sqrt{2\pi\nu^2}} \left(1 + \frac{\nu^3}{6}
\frac{\xi_R^{(3)}(0)}{\s_R^3}\right ),\label{eq:npk2}
\eea
where for the second equality we have again assumed small correlations
and truncated the series at the three-point function $\xi^{(3)}_R$.  In the
third equality, we have further used the $\nu\gg 1$ limit. 

We can now assemble the velocity distribution $\P(\ubr{1},\ubr{2};\vx_1,\vx_2)$,
collecting all non-vanishing two-point and three-point terms.  First,
we order terms in the following way:
\bea
\P(\ubr{1},\ubr{2};\vx_1,\vx_2) & = & \left\{ A + B \frac{\ubr{2}-\ubr{1}}{\s_u^2}
+ C \frac{\ubr{1}\ubr{2}}{\s_u} \right . \\
& & \  \  + \: D \left [ \frac{\ubr{1}}{\s_u}\left (\frac{\ubr{2}^2}{\s_u^2}-1\right)
- \frac{\ubr{2}}{\s_u}\left (\frac{\ubr{1}^2}{\s_u^2}-1\right) \right ]\\
& & \   \   + \left . E \left (\frac{\ubr{1}^2}{\s_u^2}-1\right)\right \} 
\frac{1}{2\pi \:\s_u^2} \exp\left(-\frac{\ubr{1}^2+\ubr{2}^2}{2\s_u^2}\right ).
\eea
The coefficients of the different velocity terms are given by:
\bea
A &=& 1 + b_L^2 \xir{\d\d}(r) + b_L^3 \left[ \frac{1}{3}\xir{\d\d\d}(0) + \xir{\d\d\d}(\vx_1,\vx_1,\vx_2) \right] - \frac{1}{3} b_L^3 \xir{\d\d\d}(0)\\
B &=& \frac{b_L}{\s_u} \xir{\d u}(r) + \frac{b_L^2}{2\s_u} \left [
\xir{\d\d u}(\vx_1,\vx_1,\vx_2) + 2 \xir{\d\d u}(\vx_1,\vx_2,\vx_2)\right ],\\
C &=& -\frac{\xir{uu}(r)}{\s_u^2} + \frac{2 b_L}{\s_u^2} \xir{\d u u}(\vx_1,\vx_1,\vx_2), \\
D &=& -\frac{1}{2 \s_u^3} \xir{uuu}(\vx_1,\vx_1,\vx_2),\\
E &=& \frac{b_L}{\s_u^2}\left[ \xir{\d uu}(0) + \xir{\d uu}(\vx_1,\vx_2,\vx_2)\right ].
\eea
Here, we have defined the first order Lagrangian bias $b_L\equiv \nu/\s_R$.  
The last term for $A$ comes from the normalization by the
number of peaks, \refeq{npk2}.  We can now evaluate \refeq{xipk2} to
obtain the peak correlation function (in real space),
\be
1 + \xi_{\rm pk}(r) = b_L^2 \xir{\d\d}(r) + b_L^3 \xir{\d\d\d}(r),
\label{eq:xipk3}
\ee
recovering the well-known expression from e.g. \cite{MV08}.  Correspondingly,
the real-space peak power spectrum reads
\be
P_{\rm pk}(k) = b_L^2 P_R(k) + b_L^3 \p^{\d\d\d}(k).
\label{eq:Ppk}
\ee
In order to obtain the matter-weighted 
velocities, we have to divide by $(1+\xi_{\rm pk})$ [\refeq{Pu}].  This
yields 
\be
A' = A - b_L^2 \xir{\d\d}(r) - b_L^3 \xir{\d\d\d}(\vx_1,\vx_1,\vx_2) = 1.
\ee
In this way, $\Pdu$ is properly normalized to second order in the
correlations.  Performing the integral in \refeq{Pu} term by term, we obtain:
\bea
\Pdu(\d u;\vx_1,\vx_2) &=& \left\{ 1 + B \frac{\d u}{\s_u} \right .
+ (E - C) \frac{1}{4}\left(\frac{\d u^2}{\s_u^2}-2 \right) \vs
& & \  \  \left . +\; D \frac{\d u}{4\s_u} \left(\frac{\d u^2}{\s_u^2} - 6\right)  \right\} \vs
& & \frac{1}{\sqrt{4\pi \:\s_u^2}} \exp\left(-\frac{\d u^2}{4\s_u^2}\right ).
\label{eq:Pdu-f}
\eea
Clearly, this expansion is equivalent to the Edgeworth expansion [\refeq{Pdu-mom}], and the
coefficients $B$ through $E$ are easily related to the moments of $\d u$.  
\newpage
Using \refeq{xilmn} again to express the various three-point
correlations, we have
\bea
\<\d u\>  &=& 2\: B\: \s_u = 2 b_L \xir{\d u}(r) + b_L^2 [\xir{\d\d u}(r) - 2 \xir{\d u \d}(r)]\\
\< \d u^2 \> &=& 2 (1 + E - C) \s_u^2 = 2[\s_u^2 + \xir{uu}(r) + b_L \xir{uu\d}(0) + b_L \xir{uu\d}(r) - 2 b_L\xir{\d uu}(r)]\\
\< \d u^3 \> &=& 12\: D\: \s_u^3 = 6\:\xir{uuu}(r)\\
\< \d u^4 \>  - 3 \< \d u^2 \>^2 &=& 0.
\eea
These moments agree precisely with those derived in \refsec{ansatz}.\\

\end{widetext}
\bibliography{NG}

\begin{thebibliography}{27}
\expandafter\ifx\csname natexlab\endcsname\relax\def\natexlab#1{#1}\fi
\expandafter\ifx\csname bibnamefont\endcsname\relax
  \def\bibnamefont#1{#1}\fi
\expandafter\ifx\csname bibfnamefont\endcsname\relax
  \def\bibfnamefont#1{#1}\fi
\expandafter\ifx\csname citenamefont\endcsname\relax
  \def\citenamefont#1{#1}\fi
\expandafter\ifx\csname url\endcsname\relax
  \def\url#1{\texttt{#1}}\fi
\expandafter\ifx\csname urlprefix\endcsname\relax\def\urlprefix{URL }\fi
\providecommand{\bibinfo}[2]{#2}
\providecommand{\eprint}[2][]{\url{#2}}

\bibitem[{\citenamefont{{Kaiser}}(1987)}]{Kaiser}
\bibinfo{author}{\bibfnamefont{N.}~\bibnamefont{{Kaiser}}},
  \bibinfo{journal}{\mnras} \textbf{\bibinfo{volume}{227}}, \bibinfo{pages}{1}
  (\bibinfo{year}{1987}).

\bibitem[{\citenamefont{{Fisher}}(1995)}]{Fisher95}
\bibinfo{author}{\bibfnamefont{K.~B.} \bibnamefont{{Fisher}}},
  \bibinfo{journal}{\apj} \textbf{\bibinfo{volume}{448}}, \bibinfo{pages}{494}
  (\bibinfo{year}{1995}), \eprint{arXiv:astro-ph/9412081}.

\bibitem[{\citenamefont{{Fry} and {Gaztanaga}}(1993)}]{FryGaztanaga}
\bibinfo{author}{\bibfnamefont{J.~N.} \bibnamefont{{Fry}}} \bibnamefont{and}
  \bibinfo{author}{\bibfnamefont{E.}~\bibnamefont{{Gaztanaga}}},
  \bibinfo{journal}{\apj} \textbf{\bibinfo{volume}{413}}, \bibinfo{pages}{447}
  (\bibinfo{year}{1993}), \eprint{arXiv:astro-ph/9302009}.

\bibitem[{\citenamefont{{Desjacques} and {Seljak}}(2010)}]{DesjacquesSeljak10}
\bibinfo{author}{\bibfnamefont{V.}~\bibnamefont{{Desjacques}}}
  \bibnamefont{and} \bibinfo{author}{\bibfnamefont{U.}~\bibnamefont{{Seljak}}},
  \bibinfo{journal}{ArXiv e-prints}  (\bibinfo{year}{2010}),
  \eprint{1003.5020}.

\bibitem[{\citenamefont{{Dalal} et~al.}(2008)\citenamefont{{Dalal}, {Dor{\'e}},
  {Huterer}, and {Shirokov}}}]{DalalEtal08}
\bibinfo{author}{\bibfnamefont{N.}~\bibnamefont{{Dalal}}},
  \bibinfo{author}{\bibfnamefont{O.}~\bibnamefont{{Dor{\'e}}}},
  \bibinfo{author}{\bibfnamefont{D.}~\bibnamefont{{Huterer}}},
  \bibnamefont{and}
  \bibinfo{author}{\bibfnamefont{A.}~\bibnamefont{{Shirokov}}},
  \bibinfo{journal}{\prd} \textbf{\bibinfo{volume}{77}},
  \bibinfo{pages}{123514} (\bibinfo{year}{2008}), \eprint{0710.4560}.

\bibitem[{\citenamefont{{Matarrese} and {Verde}}(2008)}]{MV08}
\bibinfo{author}{\bibfnamefont{S.}~\bibnamefont{{Matarrese}}} \bibnamefont{and}
  \bibinfo{author}{\bibfnamefont{L.}~\bibnamefont{{Verde}}},
  \bibinfo{journal}{\apjl} \textbf{\bibinfo{volume}{677}}, \bibinfo{pages}{L77}
  (\bibinfo{year}{2008}), \eprint{0801.4826}.

\bibitem[{\citenamefont{{Slosar} et~al.}(2008)\citenamefont{{Slosar}, {Hirata},
  {Seljak}, {Ho}, and {Padmanabhan}}}]{SlosarEtal}
\bibinfo{author}{\bibfnamefont{A.}~\bibnamefont{{Slosar}}},
  \bibinfo{author}{\bibfnamefont{C.}~\bibnamefont{{Hirata}}},
  \bibinfo{author}{\bibfnamefont{U.}~\bibnamefont{{Seljak}}},
  \bibinfo{author}{\bibfnamefont{S.}~\bibnamefont{{Ho}}}, \bibnamefont{and}
  \bibinfo{author}{\bibfnamefont{N.}~\bibnamefont{{Padmanabhan}}},
  \bibinfo{journal}{Journal of Cosmology and Astro-Particle Physics}
  \textbf{\bibinfo{volume}{8}}, \bibinfo{pages}{31} (\bibinfo{year}{2008}),
  \eprint{0805.3580}.

\bibitem[{\citenamefont{{Creminelli} et~al.}(2006)\citenamefont{{Creminelli},
  {Nicolis}, {Senatore}, {Tegmark}, and {Zaldarriaga}}}]{CreminelliEtal06}
\bibinfo{author}{\bibfnamefont{P.}~\bibnamefont{{Creminelli}}},
  \bibinfo{author}{\bibfnamefont{A.}~\bibnamefont{{Nicolis}}},
  \bibinfo{author}{\bibfnamefont{L.}~\bibnamefont{{Senatore}}},
  \bibinfo{author}{\bibfnamefont{M.}~\bibnamefont{{Tegmark}}},
  \bibnamefont{and}
  \bibinfo{author}{\bibfnamefont{M.}~\bibnamefont{{Zaldarriaga}}},
  \bibinfo{journal}{Journal of Cosmology and Astro-Particle Physics}
  \textbf{\bibinfo{volume}{5}}, \bibinfo{pages}{4} (\bibinfo{year}{2006}),
  \eprint{arXiv:astro-ph/0509029}.

\bibitem[{\citenamefont{{Meerburg} et~al.}(2009)\citenamefont{{Meerburg}, {van
  der Schaar}, and {Stefano Corasaniti}}}]{MeerburgEtal09}
\bibinfo{author}{\bibfnamefont{P.~D.} \bibnamefont{{Meerburg}}},
  \bibinfo{author}{\bibfnamefont{J.~P.} \bibnamefont{{van der Schaar}}},
  \bibnamefont{and} \bibinfo{author}{\bibfnamefont{P.}~\bibnamefont{{Stefano
  Corasaniti}}}, \bibinfo{journal}{Journal of Cosmology and Astro-Particle
  Physics} \textbf{\bibinfo{volume}{5}}, \bibinfo{pages}{18}
  (\bibinfo{year}{2009}), \eprint{0901.4044}.

\bibitem[{\citenamefont{{Catelan} and {Scherrer}}(1995)}]{CatelanScherrer}
\bibinfo{author}{\bibfnamefont{P.}~\bibnamefont{{Catelan}}} \bibnamefont{and}
  \bibinfo{author}{\bibfnamefont{R.~J.} \bibnamefont{{Scherrer}}},
  \bibinfo{journal}{\apj} \textbf{\bibinfo{volume}{445}}, \bibinfo{pages}{1}
  (\bibinfo{year}{1995}), \eprint{arXiv:astro-ph/9405011}.

\bibitem[{\citenamefont{{Desjacques}}(2008)}]{Desjacques08}
\bibinfo{author}{\bibfnamefont{V.}~\bibnamefont{{Desjacques}}},
  \bibinfo{journal}{\prd} \textbf{\bibinfo{volume}{78}},
  \bibinfo{pages}{103503} (\bibinfo{year}{2008}), \eprint{0806.0007}.

\bibitem[{\citenamefont{{Desjacques} and {Sheth}}(2010)}]{DesjacquesSheth10}
\bibinfo{author}{\bibfnamefont{V.}~\bibnamefont{{Desjacques}}}
  \bibnamefont{and} \bibinfo{author}{\bibfnamefont{R.~K.}
  \bibnamefont{{Sheth}}}, \bibinfo{journal}{\prd}
  \textbf{\bibinfo{volume}{81}}, \bibinfo{pages}{023526}
  (\bibinfo{year}{2010}), \eprint{0909.4544}.

\bibitem[{\citenamefont{{Matarrese} et~al.}(1986)\citenamefont{{Matarrese},
  {Lucchin}, and {Bonometto}}}]{MLB}
\bibinfo{author}{\bibfnamefont{S.}~\bibnamefont{{Matarrese}}},
  \bibinfo{author}{\bibfnamefont{F.}~\bibnamefont{{Lucchin}}},
  \bibnamefont{and} \bibinfo{author}{\bibfnamefont{S.~A.}
  \bibnamefont{{Bonometto}}}, \bibinfo{journal}{\apjl}
  \textbf{\bibinfo{volume}{310}}, \bibinfo{pages}{L21} (\bibinfo{year}{1986}).

\bibitem[{\citenamefont{{Grinstein} and {Wise}}(1986)}]{GrinsteinWise}
\bibinfo{author}{\bibfnamefont{B.}~\bibnamefont{{Grinstein}}} \bibnamefont{and}
  \bibinfo{author}{\bibfnamefont{M.~B.} \bibnamefont{{Wise}}},
  \bibinfo{journal}{\apj} \textbf{\bibinfo{volume}{310}}, \bibinfo{pages}{19}
  (\bibinfo{year}{1986}).

\bibitem[{\citenamefont{{Matsubara}}(2003)}]{Matsubara03}
\bibinfo{author}{\bibfnamefont{T.}~\bibnamefont{{Matsubara}}},
  \bibinfo{journal}{\apj} \textbf{\bibinfo{volume}{584}}, \bibinfo{pages}{1}
  (\bibinfo{year}{2003}).

\bibitem[{\citenamefont{{Verde} and {Matarrese}}(2009)}]{VM09}
\bibinfo{author}{\bibfnamefont{L.}~\bibnamefont{{Verde}}} \bibnamefont{and}
  \bibinfo{author}{\bibfnamefont{S.}~\bibnamefont{{Matarrese}}},
  \bibinfo{journal}{\apjl} \textbf{\bibinfo{volume}{706}}, \bibinfo{pages}{L91}
  (\bibinfo{year}{2009}), \eprint{0909.3224}.

\bibitem[{\citenamefont{{Scherrer} and
  {Bertschinger}}(1991)}]{ScherrerBertschinger}
\bibinfo{author}{\bibfnamefont{R.~J.} \bibnamefont{{Scherrer}}}
  \bibnamefont{and}
  \bibinfo{author}{\bibfnamefont{E.}~\bibnamefont{{Bertschinger}}},
  \bibinfo{journal}{\apj} \textbf{\bibinfo{volume}{381}}, \bibinfo{pages}{349}
  (\bibinfo{year}{1991}).

\bibitem[{\citenamefont{{Feldman} et~al.}(2009)\citenamefont{{Feldman},
  {Watkins}, and {Hudson}}}]{FeldmanEtal}
\bibinfo{author}{\bibfnamefont{H.~A.} \bibnamefont{{Feldman}}},
  \bibinfo{author}{\bibfnamefont{R.}~\bibnamefont{{Watkins}}},
  \bibnamefont{and} \bibinfo{author}{\bibfnamefont{M.~J.}
  \bibnamefont{{Hudson}}}, \bibinfo{journal}{ArXiv e-prints}
  (\bibinfo{year}{2009}), \eprint{0911.5516}.

\bibitem[{\citenamefont{{Kashlinsky} et~al.}(2010)\citenamefont{{Kashlinsky},
  {Atrio-Barandela}, {Ebeling}, {Edge}, and {Kocevski}}}]{KashlinskyEtal}
\bibinfo{author}{\bibfnamefont{A.}~\bibnamefont{{Kashlinsky}}},
  \bibinfo{author}{\bibfnamefont{F.}~\bibnamefont{{Atrio-Barandela}}},
  \bibinfo{author}{\bibfnamefont{H.}~\bibnamefont{{Ebeling}}},
  \bibinfo{author}{\bibfnamefont{A.}~\bibnamefont{{Edge}}}, \bibnamefont{and}
  \bibinfo{author}{\bibfnamefont{D.}~\bibnamefont{{Kocevski}}},
  \bibinfo{journal}{\apjl} \textbf{\bibinfo{volume}{712}}, \bibinfo{pages}{L81}
  (\bibinfo{year}{2010}), \eprint{0910.4958}.

\bibitem[{\citenamefont{{Lee} and {Komatsu}}(2010)}]{LeeKomatsu}
\bibinfo{author}{\bibfnamefont{J.}~\bibnamefont{{Lee}}} \bibnamefont{and}
  \bibinfo{author}{\bibfnamefont{E.}~\bibnamefont{{Komatsu}}},
  \bibinfo{journal}{ArXiv e-prints}  (\bibinfo{year}{2010}),
  \eprint{1003.0939}.

\bibitem[{\citenamefont{{Scoccimarro}}(2004)}]{Sc04}
\bibinfo{author}{\bibfnamefont{R.}~\bibnamefont{{Scoccimarro}}},
  \bibinfo{journal}{\prd} \textbf{\bibinfo{volume}{70}},
  \bibinfo{pages}{083007} (\bibinfo{year}{2004}),
  \eprint{arXiv:astro-ph/0407214}.

\bibitem[{\citenamefont{{Lam} et~al.}(2010)\citenamefont{{Lam}, {Desjacques},
  and {Sheth}}}]{LamDesjacquesSheth}
\bibinfo{author}{\bibfnamefont{T.~Y.} \bibnamefont{{Lam}}},
  \bibinfo{author}{\bibfnamefont{V.}~\bibnamefont{{Desjacques}}},
  \bibnamefont{and} \bibinfo{author}{\bibfnamefont{R.~K.}
  \bibnamefont{{Sheth}}}, \bibinfo{journal}{\mnras}
  \textbf{\bibinfo{volume}{402}}, \bibinfo{pages}{2397} (\bibinfo{year}{2010}),
  \eprint{0908.2285}.

\bibitem[{\citenamefont{{Smith} et~al.}(2003)\citenamefont{{Smith}, {Peacock},
  {Jenkins}, {White}, {Frenk}, {Pearce}, {Thomas}, {Efstathiou}, and
  {Couchman}}}]{halofit}
\bibinfo{author}{\bibfnamefont{R.~E.} \bibnamefont{{Smith}}},
  \bibinfo{author}{\bibfnamefont{J.~A.} \bibnamefont{{Peacock}}},
  \bibinfo{author}{\bibfnamefont{A.}~\bibnamefont{{Jenkins}}},
  \bibinfo{author}{\bibfnamefont{S.~D.~M.} \bibnamefont{{White}}},
  \bibinfo{author}{\bibfnamefont{C.~S.} \bibnamefont{{Frenk}}},
  \bibinfo{author}{\bibfnamefont{F.~R.} \bibnamefont{{Pearce}}},
  \bibinfo{author}{\bibfnamefont{P.~A.} \bibnamefont{{Thomas}}},
  \bibinfo{author}{\bibfnamefont{G.}~\bibnamefont{{Efstathiou}}},
  \bibnamefont{and} \bibinfo{author}{\bibfnamefont{H.~M.~P.}
  \bibnamefont{{Couchman}}}, \bibinfo{journal}{MNRAS}
  \textbf{\bibinfo{volume}{341}}, \bibinfo{pages}{1311} (\bibinfo{year}{2003}),
  \eprint{arXiv:astro-ph/0207664}.

\bibitem[{\citenamefont{{McDonald} and {Seljak}}(2009)}]{McDonaldSeljak08}
\bibinfo{author}{\bibfnamefont{P.}~\bibnamefont{{McDonald}}} \bibnamefont{and}
  \bibinfo{author}{\bibfnamefont{U.}~\bibnamefont{{Seljak}}},
  \bibinfo{journal}{Journal of Cosmology and Astro-Particle Physics}
  \textbf{\bibinfo{volume}{10}}, \bibinfo{pages}{7} (\bibinfo{year}{2009}),
  \eprint{0810.0323}.

\bibitem[{\citenamefont{{White} et~al.}(2009)\citenamefont{{White}, {Song}, and
  {Percival}}}]{WhiteSongPercival}
\bibinfo{author}{\bibfnamefont{M.}~\bibnamefont{{White}}},
  \bibinfo{author}{\bibfnamefont{Y.}~\bibnamefont{{Song}}}, \bibnamefont{and}
  \bibinfo{author}{\bibfnamefont{W.~J.} \bibnamefont{{Percival}}},
  \bibinfo{journal}{\mnras} \textbf{\bibinfo{volume}{397}},
  \bibinfo{pages}{1348} (\bibinfo{year}{2009}), \eprint{0810.1518}.

\bibitem[{\citenamefont{{Zhang} et~al.}(2007)\citenamefont{{Zhang}, {Liguori},
  {Bean}, and {Dodelson}}}]{ZhangEtal}
\bibinfo{author}{\bibfnamefont{P.}~\bibnamefont{{Zhang}}},
  \bibinfo{author}{\bibfnamefont{M.}~\bibnamefont{{Liguori}}},
  \bibinfo{author}{\bibfnamefont{R.}~\bibnamefont{{Bean}}}, \bibnamefont{and}
  \bibinfo{author}{\bibfnamefont{S.}~\bibnamefont{{Dodelson}}},
  \bibinfo{journal}{Physical Review Letters} \textbf{\bibinfo{volume}{99}},
  \bibinfo{pages}{141302} (\bibinfo{year}{2007}), \eprint{0704.1932}.

\bibitem[{\citenamefont{{Song} and {Percival}}(2009)}]{SongPercival09}
\bibinfo{author}{\bibfnamefont{Y.}~\bibnamefont{{Song}}} \bibnamefont{and}
  \bibinfo{author}{\bibfnamefont{W.~J.} \bibnamefont{{Percival}}},
  \bibinfo{journal}{Journal of Cosmology and Astro-Particle Physics}
  \textbf{\bibinfo{volume}{10}}, \bibinfo{pages}{4} (\bibinfo{year}{2009}),
  \eprint{0807.0810}.

\end{thebibliography}

\end{document}